\documentclass[twocolumn,amsmath,amssymb, prl, showpacs]{revtex4}

\usepackage{graphicx, subfigure}
\usepackage{bm}

\begin{document}

\title{Rapid laser-free ion cooling by controlled collision}

\author{Hoi-Kwan Lau\footnote{kero.lau@mail.utoronto.ca}}
\affiliation{Center for Quantum Information and Quantum Control (CQIQC),
 Department of Physics, University of Toronto, 60 Saint George Street, Toronto, M5S 1A7 Ontario, Canada}

\date{\today}

\begin{abstract}
I propose a method to remove the axial motional excitation from an ion qubit within a few oscillation periods of a harmonic trap. The principle is to prepare another coolant ion in its motional ground state, and then apply a phonon beam splitter to swap the motional state between the ions without affecting the internal state.  This method requires only the precise control of the trapping potential, thus its performance does not suffer from the limitations of current laser-cooling schemes.
With sufficient coolant ions pre-prepared, this method can rapidly re-cool ion qubits during quantum information processing.

\end{abstract}

\pacs{03.67.Lx, 37.10.Ty, 37.10.Mn}

\maketitle

\textit{Introduction.}---The ion trap system is an auspicious implementation of quantum computers (QC) \cite{Cirac:1995p2238, Blatt:2008p5155}, where various building blocks \cite{DiVincenzo:2000p14183, ANielsen:2000p5658} have been demonstrated experimentally \cite{Brown:2011p14184,Leibfried:2003p7149, Langer:2005p6869, Myerson:2008p7428}.  Achieving a high fidelity logical operation requires minimal motional excitation of the ions.  However during computation, the ions are unavoidably heated up by, for example, fluctuations of the trapping potential, imprecise ion transportation, and recoil during fluorescence readout.  
In practice, the ions have to be frequently re-cooled by sympathetic cooling \cite{Kielpinski:2000p14013}, which takes about hundreds of microseconds ($\mu s$) \cite{Jost:2009p8013}.  The cooling time has recently been improved to tens-$\mu s$ range using electromagnetically-induced-transparency techniques \cite{Morigi:2000p14417, Morigi:2003p14418, Lin:2013p14435}.  However, the cooling process remains a speed bottleneck of an ion trap QC because its duration is an order of magnitude longer than other operations \cite{Haffner:2008p3364}.  
A recent proposal suggests that $\mu s$ range cooling can be achieved by using sequences of strong laser pulses
\cite{Machnes:2010p11179}, but this method is subjected to the limitation from the laser's power.


In this letter, I describe a scheme that can rapidly re-cool a pair of ion qubit without applying laser cooling during the computation.  The scheme is divided into three processes.  Firstly, coolants, i.e. ions heterogeneous from the qubit ions which will not be involved in quantum logic gates, are prepared in the motional ground state.  Each coolant is stored in an individual harmonic well inside a segmented linear trap \cite{Wineland:1998p10591, Kielpinski:2002p7096}, with multiple controllable locally harmonic potential wells.  The second process is the core of the cooling scheme, which is a \textit{swapping beam splitter} (SBS).  As an extension of the phonon beam splitter described in \cite{Lau:2012p14011}, a SBS swaps the motional states of two ions with different masses.  
When an ion qubit has to be re-cooled, it is brought to the linear trap of a coolant.  A SBS is then applied to transfer the motional ground state from the coolant to the qubit, thus the qubit is effectively cooled.  If excessive coolants are prepared, a new coolant can be employed in each round of cooling, thus laser cooling is not required during the computation.  
The last process is to combine the individually cooled qubits to a ground state qubit pair.  This can be done by reversing the heatingless ion separation process presented in \cite{Lau:2012p14011}.  Both the SBS and the ion combination can be implemented by a controlled collision of the ions in a precisely manipulated double well trapping potential.  I will show that these processes can take less than ten trap oscillation periods, so the total process durations are at the $\mu s$ range for state-of-the-art MHz traps.  The necessary rapid and precise control of a double well potential has been demonstrated using micro-fabricated surface traps  \cite{Bowler:2012p13647}.

\begin{figure}
\begin{center}
\includegraphics{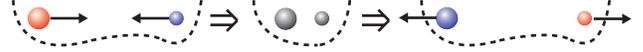}
\caption{Outline of the SBS.  Controlled collision of a qubit (large) and a coolant (small) is mediated by the variation of the double well potential (dotted).  The excited (red) and ground (blue) motional states are swapped after the process. \label{fig:outline}}
\end{center}
\end{figure}


\textit{Model}---Because linear ion transportation can be diabatic, i.e., arbitrarily fast without causing any motional excitation \cite{Torrontegui:2011p9810, Lau:2011p11390,Chen:2011p13569,Walther:2012p14012,Bowler:2012p13647}, I consider the ground state coolants are restricted to move in linear traps.  The heated qubit has to be transported to the coolant's trap for cooling.  A SBS is implemented by a controlled collision of the two ions (see Fig.~\ref{fig:outline}).  The ions are radially tightly confined but weakly trapped in the axial ($x$) direction.  Around the position of the ions, the global axial potential can be locally approximated as a double well.  The axial motional state of the ions $|\Psi\rangle$ is governed by the equation $i\hbar\partial_t |\Psi\rangle = \hat{\mathcal{H}}|\Psi\rangle$, where
\begin{equation}\label{eq:fullH}
\hat{\mathcal{H}} = \sum_{i=1}^2 \Big(\frac{\hat{P}_i^2}{2m_i}+\frac{1}{2}\xi_i^2 (t) \big(\hat{X}_i-R_i(t)\big)^2 \Big)+\frac{e^2}{4 \pi \epsilon_0 (\hat{X}_1-\hat{X}_2)}~. \nonumber
\end{equation}
The qubit (ion $1$) and the coolant (ion $2$) can be different in mass, i.e., $m_1\neq m_2$. 
The four local trap parameters, $\xi_1^2(t)$, $\xi_2^2(t)$, $R_1(t)$, and $R_2(t)$, 
 are assumed to be independently controllable by tuning the global trap potential.  The variation of these parameters will be specified by four constraints that leads to a SBS operation.

The motional excitation is characterised by the quantum fluctuation around the classical displacement of the ions.  Let us define the state of the quantum fluctuation as $|\psi\rangle \equiv \hat{D}_1^\dag(x_1,p_1)\hat{D}_2^\dag(x_2, p_2)|\Psi\rangle$, where $\hat{D}_i(x_i,p_i)=\exp\big(i(x_i\hat{P}_i-p_i\hat{X}_i)/\hbar\big)$ is the displacement operator; $x_i$ and $p_i$ are classical parameters that could be chosen as the classical position and momentum of ion $i$.  Neglecting a global phase, the state $|\psi\rangle$ obeys the equation
\begin{equation}\label{eq:H012}
i\hbar\partial_t |\psi\rangle  = (\hat{H}_1 + \hat{H}_2) |\psi\rangle~.
\end{equation}
$\hat{H}_1$ involves only the first order position and momentum operators, i.e., $\hat{H}_1 = \mathcal{V}_1\hat{p}_1 +\mathcal{V}_2\hat{p}_2 + \mathcal{F}_1\hat{q}_1+ \mathcal{F}_2\hat{q}_2$, where
\begin{equation}\label{eq:eom}
\mathcal{V}_i = \frac{p_i}{m_i}-\dot{x}_i~;~\mathcal{F}_i=\dot{p}_i+\xi_i^2(t) \left(x_i-R_i(t)\right) +\frac{(-1)^i e^2}{4 \pi \epsilon_0 r^2}~.
\end{equation}
$r\equiv x_1-x_2>0$ is the ion separation.  
The above equations reproduce the classical equation of motion when $\mathcal{V}_i=0$ and $\mathcal{F}_i=0$, thus $\hat{H}_1=0$ for my choice of $x_i$ and $p_i$.  For simplicity, \textbf{Constraint 1} will be imposed to require symmetric ion motion: local trap parameters are tuned so that $x_1=-x_2=r/2$ at any time $t$.

The dynamics of the quantum fluctuation is governed only by $\hat{H}_2$ that involves higher order terms of operators:
\begin{equation}\label{eq:quant}
\hat{H}_2= \sum_{i=1}^2\Big(\frac{\hat{p}_i^2}{2m_i}
+\frac{1}{2}\xi_i^2(t) \hat{q}_i^2\Big)
+\frac{e^2}{4 \pi \epsilon_0 r^3}(\hat{q}_1-\hat{q}_2)^2+ O(\hat{q}^3)~.
\end{equation}
For clarity, I have recast the position and momentum operators of the quantum fluctuation as $\hat{q}$ and $\hat{p}$ respectively, and from now on I refer the "motional state" to that of quantum fluctuation.  The Coulomb potential is Taylor-expanded with respect to $(\hat{q}_1-\hat{q}_2)/r$.  For the moment, quadratic approximation is applied on $\hat{H}_2$, i.e., $O(\hat{q}^3)$ is neglected.   This approximation is valid in our case because the spread of ions' wavefunction is much shorter than the ion separation, i.e., $\sqrt{\langle\hat{q}_i^2\rangle}\ll r$.

\textit{Cooling}---When two ions are brought close, their motional states will be coupled by the Coulomb interaction.  Recent experiments have employed this effect to swap the motional states between separately trapped ions \cite{Brown:2011p11810,Harlander:2011p11811}.  However, their scheme requires hundreds of $\mu$s, in order to suppress the off-resonant interactions that induce parametric excitations \footnote{The speed of the experiments is also limited by the weak Coulomb interaction between far separated ions.  The problem is claimed to be the difficulties of realising a narrowly separated double well global potential.  Such potential is not necessary in my scheme as the ions remain separated by their mutual Coulomb force when the ion separation is small.}.  In contrast, my proposal can be two orders of magnitude faster, because the parametric excitation is eliminated at construction by using the exact solution of time dependent harmonic oscillators \cite{LewisJr:1969p11779}.

Within the quadratic approximation of $\hat{H}_2$, the evolution, which is characterised by the operator $\hat{U}_t$ at time $t$, is a two-mode squeezing operation on the ions' motional state \cite{Lloyd:1999p11945, Weedbrook:2012p13102}.  The squeezing parameters depend on the tuneable local trap strength, $\xi_i^2(t)$, and on the Coulomb coupling that is determined by $r$, which is controllable by adjusting $R_i(t)$.  Our goal is to obtain the trap parameters that the two-mode squeezing becomes a SBS, i.e., after the process at $0<t<T$ the annihilation operators transform as
\begin{equation}\label{eq:swap}
\hat{U}^\dag_T \hat{a}_1\hat{U}_T=\hat{a}_2 e^{i\theta}~~;~~\hat{U}^\dag_T \hat{a}_2\hat{U}_T=\hat{a}_1 e^{i\theta}~,
\end{equation}
where $\hat{a}_i \equiv \big(\sqrt{\hbar\sqrt{m_i}\xi_i(0)}\hat{q}_i+ i\hat{p}_i/\sqrt{\hbar\sqrt{m_i}\xi_i(0)} \big)/\sqrt{2}$.  The phase factor $\theta$ does not affect the cooling performance.
The local trap strength before and after the SBS is the same, i.e., $\xi^2_i(0)=\xi^2_i(T)$, so the initial and final states can be characterised by the same $\hat{a}_i$.

To see how a SBS cools the qubit, consider the initial motional state of the coolant is the ground state while that of the qubit is an arbitrary pure state, i.e., $|\psi(0)\rangle = \int f(\alpha) |\alpha\rangle_1 d\alpha \otimes |0\rangle_2$, for some complex function $f$.  The state lies in the eigensubspace of the coolant's phonon number operator: $\hat{a}_2^\dag\hat{a}_2 |\psi(0)\rangle=0$.  According to Eq.~(\ref{eq:swap}), the SBS transforms the eigenvalue equation as $\hat{a}_1^\dag\hat{a}_1|\psi(T)\rangle=\hat{U}_T\hat{U}^\dag_T\hat{a}_1^\dag\hat{U}_T\hat{U}^\dag_T\hat{a}_1\hat{U}_T|\psi(0)\rangle=\hat{U}_T\hat{a}_2^\dag\hat{a}_2|\psi(0)\rangle=0$.  This derivation implies that the qubit will result in the ground motional state.  Since the eigenvalue equation is valid for any complex function $f$, the SBS can cool a qubit with any initial motional state.



\textit{Swapping Beam Splitter}---The construction of a SBS is clearer when considering the collective modes of the quantum fluctuation.  Let us define the quadrature operators of the centre-of-mass (+) mode and the stretching (-) mode as
\begin{equation}\label{eq:qpm}
\hat{q}_\pm \equiv \frac{1}{\sqrt{2}}\Big(\hat{q}_1\pm\sqrt{\frac{m_2}{m_1}}\hat{q}_2 \Big)~;~\hat{p}_\pm = \frac{1}{\sqrt{2}}\Big(\hat{p}_1\pm\sqrt{\frac{m_1}{m_2}}\hat{p}_2 \Big). \nonumber
\end{equation}
Then in the quadratic approximation as $\hat{H}_2$ can be re-written as
\begin{equation}\label{eq:modes}
\hat{H}_2\approx\frac{\hat{p}^2_+}{2m_1}+\frac{\hat{p}^2_-}{2m_1} +\frac{1}{2}m_1\omega_+^2(t) \hat{q}^2_+ +\frac{1}{2}m_1\omega_-^2(t) \hat{q}^2_- +\mathcal{E}\hat{q}_+\hat{q}_-~, \nonumber
\end{equation}
where the coupling strength between the modes is
\begin{equation}\label{eq:cross}
\mathcal{E}=\frac{\xi_1^2(t)}{2}-\frac{m_1}{m_2}\frac{\xi_2^2(t)}{2}+\Big(1-\frac{m_1}{m_2}\Big)\frac{e^2}{4 \pi \epsilon_0 r^3}~,
\end{equation}
and the mode frequencies are
\begin{equation}\label{eq:omegapm}
\omega_\pm^2(t) = \frac{\xi_1^2(t)}{2m_1}+\frac{\xi_2^2(t)}{2m_2}+\Big(\sqrt{\frac{1}{m_1 m_2}}\mp \frac{1}{m_1}\Big)\frac{2e^2}{4\pi\epsilon_0 r^3}~.
\end{equation}

The system can be treated as two coupled time dependent harmonic oscillators, which can be analytically solved if the + and - modes are decoupled, i.e. $\mathcal{E}=0$.  According to Eq.~(\ref{eq:cross}), this can be achieved by imposing \textbf{Constraint 2} on the coolant's local trap strength as: $\xi^2_2(t)=m_2\xi_1^2(t)/m_1+(m_2/m_1-1)e^2/2\pi\epsilon_0r^3$.



Let us define the annihilation operators of the modes as $\hat{a}_\pm \equiv \big(\sqrt{\hbar m_1 \omega_0} \hat{q}_\pm + i\hat{p}_\pm/\sqrt{\hbar m_1 \omega_0} \big)/\sqrt{2}$, 
where $\omega_0\equiv \xi_1(0)/\sqrt{m_1}$ is the qubit's initial trap frequency.  
The annihilation operators of the collective modes and the ions' motional state are related as $\hat{a}_\pm=(\hat{a}_1\pm \hat{a}_2)/\sqrt{2}$.  For a beam splitter without final parametric excitation, the modes can only be phase-shifted after the process, i.e., $\hat{U}^\dag_T\hat{a}_\pm \hat{U}_T = \hat{a}_\pm e^{-i \theta_\pm(T)}$.  
Additionally, the beam splitter is a SBS, i.e. Eq.~(\ref{eq:swap}) is satisfied, if the modes acquire a $\pi$ phase difference, i.e., $\theta_-(T) - \theta_+(T) = \pi$.

For simplicity, the + mode frequency can be required to be a constant throughout the process, i.e., $\omega_+(t)=\omega_0$.  Then + mode is not parametric excited and the phase is $\theta_+(t) = \omega_0 t$.  This can be done by, according to Eq.~(\ref{eq:omegapm}) and previous constraints, imposing the \textbf{Constraint 3} on the qubit's local trap strength as: $\xi_1^2(t) = m_1 \omega_0^2+(1-\sqrt{m_1/m_2})e^2/2\pi\epsilon_0r^3$.

According to Eq.~(\ref{eq:omegapm}) and with \textbf{Constraints 1-3} satisfied, $\omega^2_-(t)$ is uniquely determined by the ion separation:
\begin{equation}
\omega_-^2(t) = \omega_0^2 + \frac{e^2}{\sqrt{m_1m_2}\pi \epsilon_0r^3}~. \nonumber
\end{equation}
\textbf{Constraint 4} can be required on the local trap parameters so that the ion separation is appropriately varied to produce any desired $\omega_-^2(t)$.  The remaining problem is to find the time variation of $\omega_-^2(t)$ such that the - mode is not parametric excited and acquires the desired phase. 

Here I introduce an inverse engineering method to find a suitable $\omega^2_-(t)$.  As - mode is a time dependent harmonic oscillator, its evolution can be exactly solved by using the dynamic invariant formalism \cite{LewisJr:1969p11779}.  At any time $t$, the annihilation operator transform as \cite{LewisJr:1969p11779, Lau:2012p14011}
$\hat{U}^\dag_t \hat{a}_- \hat{U}_t = \eta^{(+)}(t) e^{-i\theta_-(T)} \hat{a}_- + \eta^{(-)}(t) e^{i\theta_-(T)} \hat{a}_-^\dag$,
where $\eta^{(\pm)}=(b\pm b^{-1} + i \dot{b}/\omega_0)/2$ and $\theta_-(T) \equiv \int_0^t\omega_0/b^2(t')dt'$.  These parameters are uniquely determined by a real scalar auxiliary function $b(t)$ that satisfies
\begin{equation}\label{eq:b}
\ddot{b}(t) + \omega_-^2(t) b(t)-\omega_0^2/ b^3(t)=0~.
\end{equation}

Parametric excitation is absent if $\eta^{(+)}=1$ and $\eta^{(-)}=0$.  This imposes a boundary condition (BC) on the auxiliary function: $b(t>T)\rightarrow1$.  An additional BC is required on $b(t)$ that yields the desired final phase $\theta_-(T) = \theta_+(T)+\pi= \omega_0 T + \pi$.  


In order to obtain the local trap parameters that realise a SBS, let us adopt an ansatz for $b(t)$ to
obtain an appropriate $\omega^2_-(t)$.  The ion separation, $r(t)$, is then determined by \textbf{Constraint 4}.  $\xi_1^2(t)$ and $\xi_2^2(t)$ are obtained by \textbf{Constraint 3} and \textbf{2} respectively.  Finally, $R_1(t)$ and $R_2(t)$ are obtained by \textbf{Constraint 1} and the classical equation of motion Eq.~(\ref{eq:eom}).



\textit{Ansatz}---I here suggest a class of ansatz of $b(t)$ that all the BC are satisfied at construction: 
\begin{equation}\label{eq:ansatz}
b(t)=\Big(\frac{\sqrt{\pi}}{\omega_0\sigma}e^{-(t-0.5T)^2/\sigma^2}+1\Big)^{-1/2}~.
\end{equation}
The speed of the SBS is determined by the parameter $\sigma$.  For $\omega_0^2\sigma^2=2$ and $3$, the SBS process time, $T$, is $8.3$ and $10.2$ trap oscillation periods respectively \footnote{For numerical reason, $T$ is defined at when the total phase difference is at least $10^{-4}$ deviated from $\pi$, i.e., $|\theta_-(T)-\theta_+(T)-\pi| \leq 10^{-4}$.  With this setting, the qubit possesses no more than $10^{-6}$ final motional excitation at $t=T$.}.  I note that the scaled process time, $\omega_0 T$, is independent of the ions' mass and $\omega_0$, and does not affect the quality of cooling within the quadratic approximation of $\hat{H}_2$.  Therefore, the qubit cooling time is generally in the $\mu$s range if the trap frequency is a few MHz.  

\begin{figure}
\begin{center}
\includegraphics{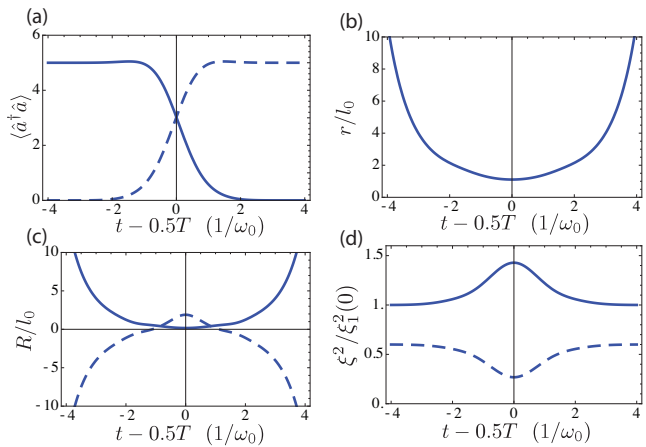}\caption{(a) Let the qubit is initially in a thermal state with $\langle \hat{a}^\dag_1 \hat{a}_1\rangle =5$.  The mean phonon number of qubit (solid) and coolant (dashed) is swapped after the SBS.  Time variation of (b) ion separation, (c) $R_1$ (solid) and $R_2$ (dashed) \cite{Rnote}, (d) $\xi_1^2$ (solid line) and $\xi_2^2$ (dashed line), for the SBS following the ansatz Eq. (\ref{eq:ansatz}) with $\omega_0^2\sigma^2=2$.   All length are expressed in terms of a characteristic length, $l_0 = \sqrt[3]{2e^2/4\pi \epsilon_0 m_1\omega_0^2}\approx 5.61 \mu$m, which is the separation of two $^{40}\textrm{Ca}^+$ ions in a single harmonic well with $\omega_0=2\pi$ MHz.
\label{fig:b}}
\end{center}
\end{figure}

As an example, a controlled collision between a $^{40}\textrm{Ca}^+$ qubit and a $^{24}\textrm{Mg}^+$ coolant with $\omega_0=2\pi$ MHz was simulated.  The cooling time is $T=1.3\mu$s for $\omega_0^2\sigma^2=2$.  Time variation of mean phonon number, ion separation, and local trap parameters are shown in Fig.~\ref{fig:b}.

\begin{figure}
\begin{center}
\includegraphics{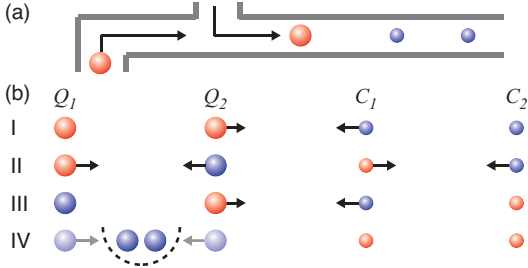}
\caption{(a) Heated qubits are transported to the linear trap containing individually trapped coolants.  (b) Sequence of constructing ground state qubit pair: (I) Motional excitation is transferred from qubit $Q_2$ to coolant $C_1$ through a SBS.  (II) Two SBS are simultaneously conducted to swap the motion between qubit $Q_1$ and $Q_2$, and between coolant $C_1$ and $C_2$.  (III) Repeat the procedure in I.  Both $Q_1$ and $Q_2$ are cooled after this step.  (IV) The qubits are combined in a single harmonic well through the heatingless ion combination process.
\label{fig:pair}}
\end{center}
\end{figure}

\textit{Ground state qubit pair}---Now I discuss how to use the above cooling scheme to prepare a ground state qubit pair for high fidelity logical operation.  Here I specifically consider the quantum computer architecture that is constituted by numerous interconnected traps \cite{Wineland:1998p10591, Kielpinski:2002p7096}, though the method is also applicable to other architectures that ions are movable in linear traps.  Consider two qubits are transported to the linear trap containing an array of individually trapped coolants (Fig.~\ref{fig:pair}a).  
The transportation may further heat up the qubits, but the excitation will  eventually be removed.
The qubits can be individually cooled by three rounds of SBS (Fig.~\ref{fig:pair}b I-III).  Between each round, the ions are transported by moving harmonic well with strength $m_1\omega_0^2$ ($m_2\omega_0^2$) for the qubit (coolant).  The ions' classical motion can be freely manipulated by precisely controlling the harmonic well, while the quantum fluctuation is unaffected \cite{Torrontegui:2011p9810, Lau:2011p11390,Chen:2011p13569,Walther:2012p14012,Bowler:2012p13647}.

After two qubits are individually cooled, they have to be combined in a single harmonic well (see Fig.~\ref{fig:pair}b IV) for the entanglement operation.  The ion combination can be made rapid and minimally heated by reversing the diabatic ion separation process in Ref. \cite{Lau:2012p14011}.  This heatingless ion combination can be viewed as half of a SBS, which the double well potential stops varying when it converges a single harmonic well.  In this process, the variation of trap parameters can be obtained by a similar procedure as that of SBS.  Here, \textbf{Constraint 1} is still required for symmetric motion.  \textbf{Constraint 2} is not necessary as the + and - modes are decoupled when both qubits have the same mass.  This constraint can be modified to require symmetric local trap strength, i.e., $\xi_1^2(t)=\xi_2^2(t)$.  \textbf{Constraint 3}, which requires constant + mode frequency, is retained so that the + mode is not parametric excited during the combination.  \textbf{Constraint 4} is still required to yield the desired - mode frequency by inverse engineering.  

For the ansatz of $b(t)$, we retain the condition $b(t<0)\rightarrow 1$ because the initially separated ions are not excited, while the crucial difference here is the BC of $b(t>T)$.  According to Eq.~(\ref{eq:b}) and the fact that $\omega_-^2= 3 \omega_0^2$ when two ions are trapped in a single harmonic well \cite{James:1998p2256, Lau:2012p14011}, $b(t>T)\rightarrow 3^{-1/4}$ should be reached after the process.  The - mode will not be parametric excited if $b(t)$ remains constant for $t>T$, thus the ion combination process would not cause motional excitation \cite{LewisJr:1969p11779}.  An example of ansatz of such $b(t)$ is
\begin{equation}
b(t)=(e^{(t-0.5T)/\sigma}+\sqrt[4]{3})^{-1} (1 - \sqrt[4]{3})+1/ \sqrt[4]{3}~. \nonumber
\end{equation}
As an illustration, if the initial separation between two $^{40}\textrm{Ca}^+$ is $r(0)=100 l_0$, they can be combined to a single well, i.e., $r(T)=l_0$, in $T=5.9/\omega_0\approx 0.94\mu$s for $\omega_0=2\pi$MHz and an ansatz with $\omega_0\sigma=0.2$.  


\textit{Preservation of quantum information}---A crucial question for the utility of this cooling scheme is the extent to which the quantum coherence of the ions' internal degrees of freedom will be degraded.  Possible detrimental effects includes the multipole interaction between the ions' internal state, which is significant if the ions are too close; and the dc Stark effect \cite{Lau:2011p11390}, which is serious if the qubit's speed is too fast.  
I consider again the collision of a $^{40}\textrm{Ca}^+$ qubit with a $^{24}\textrm{Mg}^+$ coolant.
For $\omega_0^2\sigma^2$ at the order of unity, the minimum ion separation is about 1 $l_0$.  This is the typical range of separation in ion trap experiments that the mutual multipole interaction is negligible.  
Furthermore, Fig.~\ref{fig:pair} shows that the qubit travels for tens of $\mu$m per $\mu$s.  In this range of speed the dc Stark effect is insignificant \cite{Lau:2011p11390}.  
Therefore the encoded quantum information would be preserved during a $\mu$s range SBS.  

\textit{Possible source of heating}---In practice, the local potential experienced by the ions is not purely harmonic, which would excite the motional state during the SBS.  The anharmonicity comes from both the global trap potential
and the Coulomb potential.  Particularly, the anharmonicity due to global potential can be suppressed by optimising the geometry and the potentials of the electrodes \cite{Home:2006p11926,Alonso:2013p14424}.  On the other hand, the Coulomb anharmonicity involves interaction between ions that could not be fully suppressed by adjusting the trap potential.

The effect of Coulomb anharmonicity on the performance of the cooling process was assessed numerically.  Eq.~(\ref{eq:H012}) is integrated when the lowest order term in $O(\hat{q}^3)$ is added in Eq.~(\ref{eq:quant}),  $\frac{e^2}{4\pi \epsilon_0 r^4}(\hat{q}_1-\hat{q}_2)^3$, is included.  Our result shows that even if the qubit initially has 40 phonons, the Coulomb anharmonicity finally induces no more than $10^{-3}$ phonon on the qubit.  Therefore the anharmonicity of potential is not deemed a serious threat to the performance of the SBS cooling scheme.

\textit{Conclusion}---In this letter, I propose that the axial motional excitation of an ion qubit can be removed by a controlled collision with a coolant ion.  
The process can take less than ten oscillation periods of the trap, which is at $\mu$s range for current state MHz trap.  I have outlined a procedure to obtain the time variation of the trap parameters for such process.
The cooled individual ions can then be rapidly combined into a single well for high fidelity logical operation.
If excessive coolants are prepared before the quantum computation, lengthy laser cooling is not necessary during the computation.  Thus our scheme can improve the operational speed of an ion trap quantum computer.

I note that the core of my scheme is the tuneable quadratic interaction between oscillators, so the idea could also be applied to cool systems with similar interaction and behaviour, such as polar molecules \cite{Idziaszek:2011p14180} and nanomechanical oscillators \cite{Hensinger:2005p12327, Tian:2004p14437}.

\textit{Acknowledgement}---I thank Daniel James for useful discussions.  I would like to acknowledge support from the NSERC CREATE Training Program in Nanoscience and Nanotechnology, and the E.F. Burton Fellowship.

\bibliographystyle{phaip}
\pagestyle{plain}
\bibliography{cool}

\begin{thebibliography}{10}

\bibitem{Cirac:1995p2238}
J.~I. Cirac and P.~Zoller,
\newblock Phys. Rev. Lett. {\bf 74}, 4091 (1995).

\bibitem{Blatt:2008p5155}
R.~Blatt and D.~Wineland,
\newblock Nature (London) {\bf 453}, 1008 (2008).

\bibitem{DiVincenzo:2000p14183}
D.~P. DiVincenzo,
\newblock Fortschr. Phys. {\bf 48}, 771 (2000).

\bibitem{ANielsen:2000p5658}
Michael A. Nielsen and Isaac L. Chuang, \textit{Quantum computation and quantum
  information} (Cambridge University Press, Cambridge, England, 2000).

\bibitem{Brown:2011p14184}
K.~Brown et~al.,
\newblock Phys. Rev. A {\bf 84}, 030303 (2011).

\bibitem{Leibfried:2003p7149}
D.~Leibfried et~al.,
\newblock Nature (London) {\bf 422}, 412 (2003).

\bibitem{Langer:2005p6869}
C.~Langer et~al.,
\newblock Phys. Rev. Lett. {\bf 95}, 060502 (2005).

\bibitem{Myerson:2008p7428}
A.~H. Myerson et~al.,
\newblock Phys. Rev. Lett. {\bf 100}, 200502 (2008).

\bibitem{Kielpinski:2000p14013}
D.~Kielpinski et~al.,
\newblock Phys. Rev. A {\bf 61}, 032310 (2000).

\bibitem{Jost:2009p8013}
J.~D. Jost et~al.,
\newblock Nature (London) {\bf 459}, 683 (2009).

\bibitem{Morigi:2000p14417}
G.~Morigi, J.~Eschner, and C.~Keitel,
\newblock Phys. Rev. lett. {\bf 85}, 4458 (2000).

\bibitem{Morigi:2003p14418}
G.~Morigi,
\newblock Phys. Rev. A {\bf 67}, 033402 (2003).

\bibitem{Lin:2013p14435}
Y.~Lin et~al.,
\newblock Phys. Rev. Lett. {\bf 110}, 153002 (2013).

\bibitem{Haffner:2008p3364}
H.~H{\"a}ffner, C.~F. Roos, and R.~Blatt,
\newblock Physics Reports {\bf 469}, 155 (2008).

\bibitem{Machnes:2010p11179}
S.~Machnes, M.~B. Plenio, B.~Reznik, A.~M. Steane, and A.~Retzker,
\newblock Phys. Rev. Lett. {\bf 104}, 183001 (2010).

\bibitem{Lau:2012p14011}
H.-K. Lau and D.~F.~V. James,
\newblock Phys. Rev. A {\bf 85}, 062329 (2012).

\bibitem{Bowler:2012p13647}
R.~Bowler et~al.,
\newblock Phys. Rev. Lett. {\bf 109}, 080502 (2012).

\bibitem{Torrontegui:2011p9810}
E.~Torrontegui et~al.,
\newblock Phys. Rev. A {\bf 83}, 013415 (2011).

\bibitem{Lau:2011p11390}
H.-K. Lau and D.~F.~V. James,
\newblock Phys. Rev. A {\bf 83}, 062330 (2011).

\bibitem{Chen:2011p13569}
X.~Chen, E.~Torrontegui, D.~Stefanatos, J.-S. Li, and J.~G. Muga,
\newblock Phys. Rev. A {\bf 84}, 043415 (2011).

\bibitem{Walther:2012p14012}
A.~Walther et~al.,
\newblock Phys. Rev. Lett. {\bf 109}, 080501 (2012).

\bibitem{Brown:2011p11810}
K.~R. Brown et~al.,
\newblock Nature (London) {\bf 471}, 196 (2011).

\bibitem{Harlander:2011p11811}
M.~Harlander, R.~Lechner, M.~Brownnutt, R.~Blatt, and W.~Haensel,
\newblock Nature (London) {\bf 471}, 200 (2011).

\bibitem{LewisJr:1969p11779}
H.~L. Jr and W.~Riesenfeld,
\newblock Journal of Mathematical Physics {\bf 10}, 1458 (1969).

\bibitem{Lloyd:1999p11945}
S.~Lloyd and S.~L. Braunstein,
\newblock Phys. Rev. Lett. {\bf 82}, 1784 (1999).

\bibitem{Weedbrook:2012p13102}
C.~Weedbrook et~al.,
\newblock Rev. Mod. Phys. {\bf 84}, 621 (2012).

\bibitem{Wineland:1998p10591}
D.~J. Wineland et~al.,
\newblock J. Res. Natl. Inst. Stand. Technol. {\bf 103}, 259 (1998).

\bibitem{Kielpinski:2002p7096}
D.~Kielpinski, C.~Monroe, and D.~J. Wineland,
\newblock Nature (London) {\bf 417}, 709 (2002).

\bibitem{James:1998p2256}
D.~F.~V. James,
\newblock Appl. Phys. B {\bf 66}, 181 (1998).

\bibitem{Home:2006p11926}
J.~P. Home and A.~M. Steane,
\newblock Quantum Inf. Comput. {\bf 6}, 289 (2006).

\bibitem{Alonso:2013p14424}
J.~Alonso, F.~Leupold, B.~Keitch, and J.~Home,
\newblock New Journal of Physics {\bf 15}, 023001 (2013).

\bibitem{Idziaszek:2011p14180}
Z.~Idziaszek, T.~Calarco, and P.~Zoller,
\newblock Phys. Rev. A {\bf 83}, 053413 (2011).

\bibitem{Hensinger:2005p12327}
W.~K. Hensinger et~al.,
\newblock Phys. Rev. A {\bf 72}, 041405 (2005).

\bibitem{Tian:2004p14437}
L.~Tian and P.~Zoller,
\newblock Phys. Rev. Lett. {\bf 93}, 266403 (2004).

\end{thebibliography}

\end{document}